%% file: article.v2.tex
\documentclass[journal]{IEEEtran}
\usepackage{amsmath,amssymb,bm}
\usepackage{subfigure}
\usepackage{graphicx}
\usepackage[ruled, linesnumbered]{algorithm2e}
\usepackage{wrapfig}
\usepackage{multirow}

\usepackage{cite}
\usepackage{textcomp}
\usepackage{url}
\usepackage{cleveref}
\usepackage{threeparttable}
\usepackage{booktabs} %
\usepackage{xcolor}
\usepackage{amsfonts}
\usepackage[colorinlistoftodos]{todonotes}
\usepackage{arydshln}
\usepackage{soul}

\PassOptionsToPackage{dvipsnames}{xcolor}  
\usepackage{color}
\usepackage{tikz}
\usetikzlibrary{arrows,positioning,fit,petri,shapes,backgrounds,decorations.pathmorphing,calc,shapes.misc, arrows, decorations.markings}

\newcommand{\eg}{e.\,g.,\ }
\newcommand{\ie}{i.\,e.,\ }

\newcommand{\et}{{et al.}}
\newcommand{\mb}{\mathbf}

\DeclareMathOperator*{\argmax}{arg\,max}

\definecolor{Gray}{gray}{0.9}

\begin{document}
%
\title{Deep Learning for Environmentally Robust Speech Recognition: An Overview of Recent Developments}

\author{Zixing~Zhang, J\"urgen Geiger, Jouni Pohjalainen, Amr El-Desoky Mousa, and~Bj{\"o}rn~Schuller
\thanks{Z.~Zhang, J.~Pohjalainen, A.~E.~Mousa, and B.~Schuller are with the Chair of Complex and Intelligent Systems, University of Passau, 94032 Passau, Germany (e-mail: zixing.zhang@uni-passau.de).}
\thanks{J.~Geiger is with the European Research Centre, Huawei Technologies, Dessauerstr.~10, 80992 Munich, Germany (e-mail: juergen.geiger@huawei.com).}
\thanks{B.~Schuller is also with the Department of Computing, Imperial College London, London SW7 2AZ, UK (e-mail: bjoern.schuller@imperial.ac.uk), and with audEERING GmbH, Gilching, Germany (email: bs@audeering.com).}
}

%

\maketitle

\begin{abstract}

Eliminating the negative effect of non-stationary environmental noise is a long-standing research topic for automatic speech recognition that stills remains an important challenge. 
Data-driven supervised approaches, including ones based on deep neural networks, have recently emerged as potential alternatives to traditional unsupervised approaches and with sufficient training, can alleviate the shortcomings of the unsupervised methods in various real-life acoustic environments. 
In this light, we review recently developed, representative deep learning approaches for tackling non-stationary additive and convolutional degradation of speech with the aim of providing guidelines for those involved in the development of environmentally robust speech recognition systems. 
We separately discuss single- and multi-channel techniques developed for the front-end and back-end of speech recognition systems, as well as joint front-end and back-end training frameworks.
\end{abstract}

\begin{IEEEkeywords}
deep learning, neural networks, robust speech recognition, non-stationary noise, multi-channel speech denoising
\end{IEEEkeywords}

\IEEEpeerreviewmaketitle

\section{Introduction}
\label{sec:introduction}

Recently, following years of research, Automatic Speech Recognition (ASR) has achieved major breakthroughs and greatly improved performance~\cite{GeorgeSaon16-IBM,Amodei16-Deep,Xiong16-Achieving}. Plenty of speech-specific intelligent human--machine communication systems, such as smartphone assistants (e.g. Siri, Cortana, Google Now), Amazon Echo, and Kinect Xbox One, have started to become part of our daily life. However, one of the central issues that limits their performance in everyday situations is still their performance degradation due to ambient noise and reverberation that corrupt the speech as captured by microphones.


According to the spectral distribution, the noises (including reverberation) can be generally grouped into stationary noise (constant with respect to time) or non-stationary noise (\ie varying with time, such as transient sound events, interfering speakers, and music).
Provided that it is possible to reliably detect instants of the absence of the target signal (\ie the speech signal of interest), short-term stationary additive noise can be adequately tackled with standard, unsupervised noise reduction signal processing techniques mainly developed in the 1970s and 1980s~\cite{Loizou13-Speech}.  
However, detecting and reducing the effects of non-stationary ambient noise, competing non-stationary sound sources, or highly reverberant environments, is still very challenging in practice~\cite{Yoshioka12-Making,Yang12-Machine,Barker13-PASCAL,Barker15-third,Kinoshita16-summary,Vincent16-analysis}. To address these issues, a new wave of research efforts has emerged over the past five years, as showcased in the robust speech recognition challenges such as REVERB and CHiME~\cite{Barker13-PASCAL,Vincent13-second,Barker15-third,Kinoshita16-summary}. 

In this research, {\em data-driven} approaches based on a supervised machine learning paradigm have received increasing attention, and have emerged as viable methods for enhancing robustness of ASR systems~\cite{Maas12-Recurrent}. 
The primary objective of these approaches is, by means of learning from large amounts of training data, to either obtain cleaner signals and features from noisy speech audio, or directly perform recognition of noisy speech.  
To this end, \emph{deep learning}, which is mainly based on \emph{deep neural networks}, has had a central role in the recent developments~\cite{Woellmer09-Improving,Woellmer10-Combining,Geiger14-Memory,Qian16-Very}. 
Deep learning has been consistently found to be a powerful learning approach in exploiting large-scale training data to build complex and dedicated analysis systems~\cite{Zhang17-Advanced}, and has achieved considerable success in a variety of fields, such as gaming~\cite{Mnih15-Human}, visual recognition~\cite{Russakovsky15-Imagenet, Liu17-Implicit}, language translation~\cite{Wu16-Googles}, music information retrieval~\cite{Schedl16-Introduction}, and ASR~\cite{Dahl12-Context,Hinton12-Deep}. These achievements have encouraged increasing research efforts on deep learning with the goal of improving the robustness of ASR in noisy environments. 

\begin{figure}[!t]
 \centering
 \input{ASRFramework.tex}
 \caption{General framework of a speech recognition system divided into front-end and back-end.}
 \label{fig:ASRframework}
\end{figure}
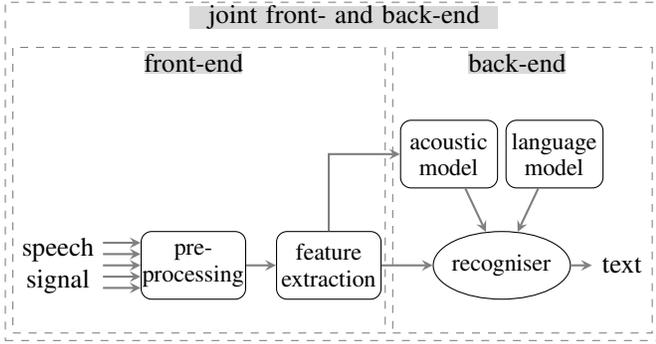

In this survey, we provide a systematic overview of relevant deep learning approaches that are designed to address the noise robustness problem for speech recognition. 
Rather than enumerating all related approaches, we aim to establish a taxonomy of the most promising approaches, which are categorised by two principles: 
i) according to the addressed number of channels, these approaches can be grouped into {\em single-channel} or {\em multi-channel} techniques; ii) according to the processing stages of an ASR system, in which deep learning methods are applied, these approaches can be generally classified into {\em front-end}, {\em back-end}, or {\em joint front- and back-end} techniques (as shown in Fig.~\ref{fig:ASRframework}). 
We highlight the advantages and disadvantages of the different approaches and paradigms, and establish interrelations and differences among the prominent techniques.
This overview assumes that the readers have background knowledge in noise-robust ASR and deep learning. However, we provide some key concepts of the raised noise-robust speech recognition problem and neural networks, \eg fully-connected layers, convolutional layers, and recurrent layers, for a better overview. For more detailed knowledge of noise-robust ASR systems or deep learning, the readers can refer to~\cite{Li14-Overview} and~\cite{Goodfellow16-Deep}, respectively. Note that, in this overview, the term \textit{deep neural networks} refers to networks including multiple hidden layers. 

Whilst several related surveys on environmentally robust speech recognition are available in the literature (\eg~\cite{Gong95-Speech,Deng11-Front,Acero12-Acoustical,Virtanen12-Techniques,Yoshioka12-Making,Li14-Overview}), none of these works focuses on the usage of deep learning. The emergence of deep learning is, however, deemed as one of the most significant advances in the field of speech recognition in the past decade and thus merits a dedicated survey.

The remainder of this article is organised as follows. In Section~\ref{sec:background}, we briefly introduce the background of this overview.
In Sections~\ref{sec:front-end} to~\ref{sec:end-to-end}, we comprehensively summarise the representative single-channel algorithms at the front-end, the back-end, and the joint front- and back-end of speech recognition systems, respectively. 
In Section~\ref{sec:multi-channel}, we then review promising multi-channel algorithms, before drawing our conclusions in Section~\ref{sec:conclusion}. 

\section{Background} 
\label{sec:background} 

In this section, we briefly describe the environmental noise problem for speech recognition. We then analyse the drawbacks and limitations of traditional approaches, and introduce the opportunities for deep learning. Finally, we introduce some standard noisy speech databases and evaluation metrics for performance comparison of the following reviewed deep learning approaches. 

\subsection{Problem Description} 
\label{subsec:problem}

In real-life scenario, the \textit{raw} speech signal $s(t)$ is easily corrupted by convolutional noise $r(t)$ (or Room Impulse Response [RIR]) and additive
noise $a(t)$ when transmitting through spatial channel. Thus, the observed distant-talk signal $y(t)$ at the microphone can be written as:
  \begin{equation} \label{eq:1}
     {y}(t)=s(t) \ast r(t) + n(t). 
  \end{equation}

When applying Short-Time Discrete Fourier Transform (STDFT) on the mixed/noisy speech, the length of RIR $T_{60}$ should be considered. If it is much shorter than the analysis window size $T$, $r(t)$ only effects the speech signals within a frame (analysis window). For many applications (\eg occurring in typical office and home environment), however, the reverberation time $T_{60}$ ranges from 200 to 1\,000\,$ms$ that is much longer than the analysis window size, resulting in an undesirable influence on the following speech frames.
For example, if the duration of a RIR is 1\,$s$ ($T_{60}$) and a feature frame is extracted at every 10\,$ms$, one RIR would smear across the following 100 frames. 
Therefore, this distorted speech in the amplitude \textit{spectral} domain, can be formulated by (see~\cite{Avargel07-System} for more details): 
\begin{equation} \label{eq:6}
  {Y}(n,f)\approx\sum_{d=0}^{D-1}S(n-d,f)R(d,f) + A(n,f),
\end{equation}
with an assumption that $r(t)$ is a constant function. Particularly, $R(d,f)$ denotes the part of $R(f)$ (\ie STDFT of RIR $r(t)$) corresponding to a frame delay $d$. In this case, the channel distortion is no longer of multiplicative nature in a linear spectral domain -- rather it is non-linear. 

Assuming that the phases of different frames, and the speech and noise signals, are non-correlated for simplification (not the case in practise), the \textit{power spectrum} of Eq.~\ref{eq:6} can be approximated as 
\begin{equation} \label{eq:7}
  \begin{aligned}
  |Y(n,f)|^2\approx\sum_{d=0}^{D-1}|S(n-d,f)|^2|R(d,f)|^2 + A^2(n,f).
  \end{aligned}
\end{equation}

Then, the following relation is obtained in the \textit{Mel spectral} domain for the $k$-th Mel-filter-bank output
\begin{equation} \label{eq:7}
  \begin{aligned}
  Y^{mel}(n,k)\approx\sum_{d=0}^{D-1}S^{mel}(n-d,k)R^{mel}(d,k) + A^{mel}(n,k), 
  \end{aligned}
\end{equation}
where $S^{mel}(n,k) = \mb{B}[k]\cdot S^2(n,f)$ with $\mb{B}=(b_{k,f})\in\mathbb{R}^{K\times F}$, $K$ is the number of Mel bins and $b_{k,f}$ is the weight of the DFT bin $f$ in the $k$-th Mel bin. $R^{mel}(n,k)$ and $A^{mel}(n,k)$ are defined similar to $S^{mel}(n,k)$. 

To extract the Mel Frequency Cepstral Coefficients (MFCCs) in \textit{cepstral} domain for ASR, logarithms and Discrete Cosine Transform (DCT) are further executed over the above mel spectral signals, so that  

\begin{equation} \label{eq:8}
   \begin{aligned}
    Y^{dct}(n,i) \approx S^{dct}(n,i) + R^{dct}(0,i) + M^{dct}(n,i), 
   \end{aligned}
\end{equation}
where $S^{dct}(n,i) = \mb{C}[i]\text{log}(S^{mel}(n,k))$ with $\mb{C}$ denoting a discrete cosine transformation matrix (same definition is for $R^{dct}(0,i)$ and $M^{dct}(n,i)$), and 
\begin{equation} \label{eq:9}
  \begin{aligned}
  M(n,i) = 1 + \frac{\sum_{d=1}^{D-1}S^{mel}(n-d,k)R^{mel}(d,k) + A^{mel}(n,k)}{S^{mel}(n,k)R^{mel}(0,k)}. 
  \end{aligned}
\end{equation}

From Eq.~(\ref{eq:1}) to (\ref{eq:8}), it can be found that the clean speech and the mixed/noisy speech have a highly \textit{non-linear} correlation in either  temporal, spectral, power spectral, mel spectral, log mel spectral, or cepstral domains, which results in a difficulty for noise cancellation.

Furthermore, the \textit{time-variant} characters of RIR and additive noise (time-invariant additive noise is beyond the scope of this paper) make the task even more challenging. For RIR, many factors can lead to a change, for instance, the position of the speaker (\ie the distance and angle between the speaker and microphone), the size, shape, and material of acoustic enclosure (such as a living room). 
For additive noise, it could be abrupt sound like thunder and bark, side talking, and also music and driving noise. All these noises are almost unpredictable.

\subsection{Deep Learning vs Traditional Approaches}
\label{subsec:traditional}

The ultimate goal of robust ASR systems is to learn well the relationship between noisy speech and the word predictions, \ie 
\begin{equation}
w=f(\mb{y}),
\end{equation}
where $\mb{y}$ denotes the representation of noisy speech $y(t)$, and $w$ is the target word. To simplify this process, we often divide it into two steps conducted at the system front end and back end, respectively. 
At the front end, speech enhancement (aka speech separation) or feature enhancement is applied to improve the quality and intelligibility of the estimated target speech on either signal level or feature level, so as to obtain the signals as clean as possible. That is, 
\begin{equation}
s(t)\leftarrow\hat{s}(t) =f_s(y(t)). 
\end{equation}
At the back end, model updating is applied to make acoustic models adapt to the new data, \ie 
\begin{equation}
w = f_m(\mb{\hat{x}}), 
\end{equation}
where $\hat{\mb{x}}$ indicates the representation from enhance speech or the enhanced representation. 

Traditional solutions on the front-end are mainly dominated by unsupervised signal processing approaches over the past several decades. {\em Spectral subtraction} \cite{Boll79-Suppression} subtracts an averaged noise spectrum (magnitude or power spectrum) from the noisy signal spectrum, while keeping the resultant spectral magnitudes positive. It only affects the spectrum magnitudes, while the spectrum phases are obtained from the noisy signal.  {\em Wiener filtering} \cite{Loizou13-Speech} adopts stochastic models and is often implemented in practice using iterative approaches which base new estimates of the filter on the enhanced signal obtained by the previous iteration's estimate \cite{Hansen91-Constrained}. 
Another popular family of techniques comprises the {\em Minimum Mean Square Error (MMSE)} \cite{Ephraim84-Speech} and {\em log-spectral amplitude MMSE (Log-MMSE)} Short-Time Spectral Amplitude (STSA) estimators \cite{Ephraim85-Speech}. Despite that they are able to yield lower musical noise, a trade-off in reducing speech distortion and residual noise needs to be made due to the sophisticated statistical properties of the interactions between speech and noise signals~\cite{Xu15-Regression}.

Most of these unsupervised methods are based on either the additive nature of the background noise, or the statistical properties of the speech and noise signals. However, they often fail to track non-stationary noise in real-world scenarios in unexpected acoustic conditions~\cite{Xu15-Regression}. Althrough some supervised machine learning approaches have been proposed, such as {\em Non-negative Matrix Factorisation} (NMF)~\cite{Lee99-Learning,Schuller10-Non,Weninger12-Supervised,Geiger14-Investigating}, they struggle to obtain effective representations (aka dictionaries) of noise and speech in complex and noisy acoustic environments.

Deep learning that is mainly based on Deep Neural Networks (DNNs), however, is well suited to address such a complex \textit{non-linear} problem~\cite{Goodfellow16-Deep}. The neural node, a basic unit constituting a network, is analogous to a biological neuron. The value of a node is usually computed as a weighted sum of the inputs followed by a non-linear activation function. Theoretically, a single node can represent a huge amount of information as long as the numerical resolution allows. Practically, deep neural networks implement multiple neural network layers (each layer consists of multiple nodes). As a result, when combining many non-linear activation functions, it enables 
the network to learn complicated relationships between the inputs and outputs. 

More specifically, typical neural layers frequently employed in deep learning include \textit{fully-connected} layer, \textit{convolutional} layer, and \textit{recurrent} layer. \textit{Fully-connected} layer is also know as dense layer and Multi-Layer Perception (MLP). 
In speech processing, stacking fully-connected layers on the top of extracted features (\eg spectrogram) has already shown a great potential to extract high-level representation for speech recognition~\cite{Dahl12-Context} via a greedy layer-wise unsupervised pre-training strategy~\cite{Hinton06-Reducing}. 

\textit{Convolutional} layer is a biologically inspired variant of fully-connected layer originally developed for visual perception tasks~\cite{LeCun89-Backpropagation}, and is the elementary layer to construct Convolutional Neural Networks (CNN). 
It employs a small size of 2D convolutional kernel `sweep' over a 2D input, and delivers a representation of local activations of patterns. 
In image processing, convolutional layer has been frequently and clearly visualised to effectively extract the hierarchical features (see~\cite{Zeiler14-Visualizing} for more details). This strongly encourages its applications to the speech domain, since the time-frequency representation of acoustic signals can be considered as an image. Besides, the 2D kernel can be modified into 1D kernel and directly applied to raw signals. Recent work has shown that the convolutional layer can automatically learn fundamental frequencies from raw signals~\cite{Trigeorgis16-Adieu}. 

In contrast to the aforementioned feed-forward layers (\ie fully-connected layer and convolutional layer), \textit{recurrent} layer (elementary layer for Recurrent Neural Networks [RNNs]) allows cyclical connections. These connections consequently endow the networks with the capability of accessing previously processed inputs. However, it cannot access long-term temporal contextual information since it suffers from the vanishing gradient problem when training. To overcome this limitation, Long Short-Term Memory (LSTM)~\cite{Hochreiter97-Long} unit and most recently Gated Recurrent Unit (GRU)~\cite{Cho14-properties} were introduced, which make the recurrent layer a powerful tool for speech analysis owing to the highly time-varying character of speech and noise. 

All these layer types, especially their stacked layers, provide deep neural networks the ability to deal with the raised problem of reducing noise and reverberation at the front end. 

At the system back end, the Hidden Markov Models (HMMs) and Gaussian Mixture Models (GMMs) were widely used as acoustic models to characterise the distribution of speech a few years ago. 
The most common ways include Maximum A Posterior (MAP)~\cite{Gauvain94-Maximum} estimation and Maximum Likelihood Linear Regression (MLLR)~\cite{Leggetter95-Maximum}. These techniques have been successfully applied to noise adaptation. 
In this article, we cannot enumerate all traditional approaches, which are beyond the scope of this survey. A systematic overview of traditional approaches on the back end can be found in~\cite{Li14-Overview}. 

In spite of the success, most of these approaches suffer the significant drawbacks: (i) they are particularly designed for generative models (\eg GMM-HMM); (ii) they assume that the adapted data match with the observed data, which is often not true in practise; (iii) they fail in  modelling large-scale data and complex environments. 

In recent years, the acoustic model has shifted from generative GMM to discriminative DNN owing to its powerful capability of representation learning. In this case, traditional approaches such as  MAP do not work any more. New noise adaptation techniques for the DNN acoustic models need to be investigated. Besides, with the rise of big data era, it is now feasible to collect huge amounts of realistic noisy speech via the microphones that are pervasive in the world. Moreover, the advance of cloud computing makes it possible that the DNN acoustic model with millions of trainable parameters can be learnt from massive noisy data.

\subsection{Standard Corpora and Evaluation Metrics}
\label{subsec:corpora}

To better compare the effectiveness of various deep learning approaches for noise-robust ASR, we introduce a set of widely used standard databases (see~Table\ref{tab:database}) in the ASR community. Among them, the earliest and most famous databases are the Aurora series developed by the European Telecommunications Standards Institute (ETSI). 
  
Note that all Aurora databases were artificially simulated, except Aurora 3~\cite{Moreno00-SPEECHDAT} that was recorded in a real noisy-car environment. All Aurora databases were created based on the clean and small-vocabulary database TIDigits for digit recognition, except Aurora 4~\cite{Pearce02-Aurora} that was constructed by corrupting Wall Street Journal (WSJ0) corpus and designed for Large Vocabulary Continuous Speech Recognition (LVCSR). All Aurora databases were mainly corrupted by additive noise, except Aurora 5~\cite{Hirsch07-Aurora} that was developed for hand-free speech recognition and simulated by involving RIR obtained in rooms and cars. 

Apart from the Aurora databases, more recently developed databases relate to CHiME series. All these CHiME databases (from the 1st to 4th) involve not only additive noise by adding various ambient noises, but also convolutional noise. More specifically, the 1st and 2nd CHiME databases~\cite{Barker13-PASCAL,Vincent13-second} include two tracks: one is for small vocabulary digit recognition based on Grid database, and the other is for LVCSR based on WSJ0; whereas the 3rd and 4th CHiME databases~\cite{Barker15-third,Vincent16-analysis} only include the data for LVCSR. Moreover, the 3rd and 4th CHiME databases considered more realistic noisy speech for evaluation, and applied a microphone array to obtain multi-channel signals. 

Other frequently used databases include REVERB~\cite{Kinoshita16-summary}, AMI~\cite{Carletta05-AMI}, and Voice Search~\cite{Schalkwyk10-Your}. Particularly, the AMI and Voice Search contain hundreds of recordings of spontaneous speech in real-life scenarios. 

Overall, the standard databases were developed for scenarios from small to large vocabulary, from artificial simulation to realistic recording, from additive noise only to convolutional noise extended, and from single channel to multiple ones. All these development trends enable the ASR systems to approach a more realistic application scenario in the wild. 

\begin{table*}
\caption{General description of some standard evaluation corpora for environmentally robust speech recognition. These corpora are either \textit{real}istically recorded or artificially \textit{sim}ulated based on certain clean databases. \textit{Add}itive and/or \textit{con}volutional noises are collected in various environments.}
\centering
\begin{threeparttable}
\begin{tabular}{llllll}
\toprule
Dataset&based on&environments&sim./real&noise types&channels \\ 
\midrule
Aurora-2~\cite{Pearce00-aurora}&TIDigits&eight conditions& sim. & add. (mainly) & single\\
Aurora-3~\cite{Moreno00-SPEECHDAT}&TIDigits&car&real & add. (mainly) & single\\
Aurora-4~\cite{Pearce02-Aurora}&WSJ0&str/tra/car/bab/res/air& sim. & add. (mainly) & dual\\
Aurora-5~\cite{Hirsch07-Aurora}&TIDigits&rooms and cars&sim. & add. \& con.  &  single\\
CHiME-1~\cite{Barker13-PASCAL}&Grid, WSJ0&home&sim& add. \& con. & dual \\ 
CHiME-2~\cite{Vincent13-second}&Grid, WSJ0&home&sim& add. \& con. & dual \\ 
CHiME-3~\cite{Barker15-third}&WSJ0&bth/bus/caf/ped/str&real \& sim &add. \& con.& six \\ 
CHiME-4~\cite{Vincent16-analysis}&WSJ0&bth/bus/caf/ped/str&real \& sim &add. \& con. & six \\
REVERB~\cite{Kinoshita16-summary} &WSJCAM0&ambient noise&real \& sim &add. \& con. & eight \\ 
AMI~\cite{Carletta05-AMI} &-&meeting&real& con. (mainly) & four/eight \\
Voice Search~\cite{Schalkwyk10-Your} &-&voice search&sim.&add. \& con. & dual \\ 
\bottomrule
\end{tabular}
\end{threeparttable}
\label{tab:database}
\end{table*}

The {\em de facto} standard metric to evaluate the performance of ASR systems is \textit{Word Error Rate} (WER) or \textit{Word Accuracy Rate} (WAR). However, to measure the performance of the front-end techniques such as speech enhancement, other intermediate subjective and objective metrics are also frequently employed. 
Specifically, typical objective metrics include \emph{segmental Signal-to-Noise Ratio} (segSNR)~\cite{Quackenbush88-Objective, Hansen98-effective}, \emph{distance measures}, \emph{Source-to-Distortion Ratio} (SDR)~\cite{Vincent06-Performance}, and \emph{Perceptual Evaluation of Speech Quality} (PESQ) \cite{ITU-862}. More detailed definitions and explanations of these objective metrics can be found in~\cite{Hu08-Evaluation}. 
Although no research has proved that a good value of these intermediate metrics for enhancement techniques necessarily leads to a better WER or WAR, experimental results have frequently shown a strong correlation between them. For example, in~\cite{Weninger15-Speech} the authors conducted speech recognition on the enhanced speech, and found that SDR and WER improvements are significantly correlated with Spearman's rho = 0.84 in single-channel case, and Spearman's rho = 0.92 in two-channel case, evaluated on the CHiME-2 benchmark database.  

\section{Front-End Techniques} 
\label{sec:front-end}

The techniques at the front end often relate to speech enhancement, source separation, and feature enhancement. 
Both \emph{speech enhancement} and \emph{source separation} attempt to obtain the estimated temporal signals as clean as possible, which can certainly be used for any speech applications including ASR. {\em Feature enhancement}, however, mainly focuses on purifying the derived features, such as MFCCs, which are largely designed for specific intelligent tasks (\ie ASR here). 
In this overview, we treat all three techniques as {\em enhancement} techniques, as they often share the same or similar algorithms. 

When applying deep learning approaches to the environmentally robust speech recognition systems, it is particularly important to effectively and efficiently represent the information of speech signals, since training DNNs is computationally intensive. In many cases, two-dimensional representations provide speech data in an effective form, and can be obtained by applying a series of operations to the raw signals $y(t)$, including Short-Time Fourier Transform (STFT, $Y(n)$), square magnitude ($|Y(n)|^2$), Mel-frequency filterbank ($Y^{mel}(n)$), log Mel-frequency filterbank ($Y^{logMel}(n)$), and even Discrete Cosine Transform (DCT, $Y^{dct}(n)$) (see Section~\ref{subsec:problem} for more details). For a better introduction of related approaches, we separately term the data spaces after each operation as {\em temporal}, {\em magnitude-spectral}, {\em power-spectral}, {\em mel-spectral}, {\em log-Mel-spectral}, and {\em Mel-cepstral} domains. Enhancement techniques can theoretically be applied to each domain, \ie from the raw signals in the temporal domain to the MFCCs in the cepstral domain.

Deep learning-based front-end techniques are normally designed in a supervised manner. For a better review, we set the input of a learning model as $\mb{y}$ that is the representation extracted from noisy speech, and the target as $\mb{x}$. 
Based on how the training target $\mb{x}$ is obtained, the techniques can be categorised into i) {\em mapping-based} methods, where $\mb{x}$ is the representation, straightforwardly extracted from clean speech, or ii) {\em masking-based} methods, where $\mb{x}$ is a mask calculated between clean and noisy speech.

\subsection{Mapping-based Deep Enhancement Methods} 
\label{subsubsec:mapping-based}

The mapping-based methods aim to learn a non-linear mapping function $F$ from the observed noisy speech $y(t)$ into the desired clean speech $s(t)$, as 
\begin{equation}
y(t)\xrightarrow{F} s(t). 
\end{equation}
Owing to the  fast-variation problems of raw speech signals and the high computational complexity they require, such a learning strategy is often applied to the data in the spectral and cepstral domains rather than the temporal domain.

To learn $F$, the neural networks are trained to reconstruct the target features $\mb{x}$ (extracted from the clean speech $s(t)$) from the corresponding input features $\mb{y}$ (extracted from the corrupted speech $y(t)$). The parameters of neural networks (models) $\theta$ are determined by minimising the objective function of the Mean Squared Error (MSE):  
\begin{equation}\label{eq:mse}
  \mathcal{J(\theta)}=\frac{1}{N}\sum_{n=1}^{N}\lVert F(\mb{y}_{n})-\mb{x}_{n}\rVert^2,
\end{equation}
where $\lVert\cdot\rVert^2$ is the squared loss, and $n$ denotes the frame index. After the estimated clean features $\hat{\mb{x}_{n}}= F(\mb{y}_{n})$ are obtained, they will be then reversed back to the time-domain signals $\hat{s}(t)$ by using the phase information from the original noisy speech, and evaluated by the objective measures as aforementioned. 

\subsubsection{Based on Stacked AutoEncoder or Deep Bolzmann Machine}
Specifically, in 2013, a Stacked AutoEncoder (SAE) was employed to map noisy speech to clean speech in the Mel-spectral domain~\cite{Lu13-Speech}.
Given an AutoEncoder (AE) that includes one non-linear encoding stage and one linear decoding stage for real valued speech as
\begin{equation}
\begin{aligned}
 h(\mb{y}) &=g(\mb{W}_1\mb{y}+\mb{b}) \\
 \mb{\hat{x}}&=\mb{W}_2h(\mb{y})+\mb{b}, 
\end{aligned}
\end{equation}
where $\mb{W}_1$ and $\mb{W}_2$ are the weight matrices of encoding and decoding, $\mb{b}$ is the bias, and $g$ denotes the activation function. 
The training pair for the first AE is $\mb{y}$ and $\mb{x}$, and then the training pair for the next AE will be $h(\mb{y})$ and $h(\mb{x})$ if weight matrices of the encoder and decoder are tied, \ie $\mb{W}_1=\mb{W}_2^T=\mb{W}$. The empirical results indicate that SAE-based enhancement methods notably outperform the traditional methods like MMSE for enhancing speech distorted by factory and car noises~\cite{Lu13-Speech}. 

Analogous to this, another successful work has been shown in~\cite{Xu14-Experimental}, where a Deep Bolzmann Machine (DBM) was utilised to estimate the complex mapping function. In the pre-training stage, noisy speech was used to train  Restricted Bolzmann Machines (RBMs) layer-by-layer in a standard unsupervised greedy fashion to obtain a deep generative model~\cite{Hinton06-Reducing}; whereas, in the fine-tuning process, the desired clean speech was set as  the target by minimising the objective function as Eq.~(\ref{eq:mse}). Similar research efforts were also extensively made on the log magnitude~\cite{Han15-Learning} and the log-Mel-spectral domains~\cite{Xu15-Regression}, respectively. 

Motivated by the fact that the same distortion in different frequency bands has different effects on speech quality, a weighted SAE was proposed in~\cite{Xia13-Speech} and showed positive performance for denoising. In detail, a weighted reconstruction loss function is employed to train SAE on the power spectrum as 
\begin{equation}
  \mathcal{J(\theta)}=\frac{1}{N}\sum_{n=1}^{N}\mb{\lambda}_w\lVert F(\mb{y}_{n})-\mb{x}_{n}\rVert^2,
\end{equation}
where $\lambda_w$ is a weight for the $w$-th frequency band.  

Further, related approaches were also shown in \cite{Ishii13-Reverberant} and \cite{Feng14-Speech}, where the authors utilised Stacked Denoising AutoEncoders (SDAEs) to enhance the Mel filterbank features corrupted by either additive or convolutional noise for ASR. The networks were pre-trained with multi-condition data, and fine-tuned by mapping the noisy speech to the clean speech. Experimental results indicate that the SDAE-based mapping method remarkably outperforms the spectral subtraction method in ASR.

\subsubsection{Based on LSTM-RNN}
For sequence-based pattern recognition, context information is considered to be vitally important~\cite{Hochreiter97-Long}. However, the aforementioned denoising networks (\ie SAE, DBM, and SDAE)  are considered to be less capable in this respect, although certain naive solutions for context-dependent processing have been applied, such as expanding several sequential frames as a long vector input~\cite{Xu14-Experimental}. 
RNNs, especially the LSTM-RNNs, have been frequently demonstrated to be highly capable of capturing the context information in a long sequence~\cite{Wollmer10-Combining,Graves13-Generating}. 

In this light, Maas \et~\cite{Maas12-Recurrent} introduced RNNs to clean distorted input features (\ie MFCCs). Specifically, the model was trained to predict clean features when presented with a noisy input frame by frame. This enhancement model has been shown to be competitive with other DNN-based mapping models at various levels of SNR when evaluated by ASR systems.  
Following from this work, W\"ollmer \et~\cite{Woellmer13-Feature} further proposed to use LSTM-RNNs to handle highly non-stationary additive noise, which was then extended to coping with reverberation in~\cite{Weninger13-Munich,Zhang14-Channel,Weninger14-Feature,Weninger14-MERL,Zhang16-Facing}. 
With the help of LSTM-RNN, the speech recognition systems perform much better than the ones without LSTM-RNN when decoding noisy speech~\cite{Weninger13-Munich,Zhang14-Channel,Weninger14-Feature,Weninger14-MERL}.

\subsubsection{Based on CNN}
Owing to the capability to capture the inherent representations embedded in the spectro-temporal feature space or in the raw signals, CNNs have attracted increasing interest in recent years~\cite{Sainath15-Convolutional,Amodei16-Deep}. For image restoration and further image processing tasks, {\em deep convolutional encoder-decoder} networks were proposed in~\cite{Mao16-Image} and delivered promising performance.
This network was further introduced for speech enhancement~\cite{Park16-Fully}, where the time-frequency spectrum (spectrogram) is viewed as an image. Specifically, the encoder network includes multiple convolutional layers in order to discover the primary information from the spectrum, and the decoder network is composed of a hierarchy of decoders, one corresponding to each encoder, for compensating the details. In order to have suitable error back-propagation to the bottom layers and to pass important information to the top layers, symmetrical links between convolutional and de-convolutional layers are added by employing skip-layer connections.

However, one main drawback of the widely used spectral or cepstral representations is discarding of potentially valuable information, such as phase. When recovering the speech, the noisy phase spectrum is straightforwardly applied in constructing the enhanced speech, even though it may suffer from distortion. 

Most recently, a novel network structure, namely WaveNet~\cite{Oord16-Wavenet}, was announced to synthesise natural speech. It takes a series of small causal and dilated convolutional layers with exponentially increasing dilation factors, which contributes to 
a receptive field growth that is exponential with respect to depth, and a significant reduction of the computational complexity. This provides an opportunity to directly map the noisy speech to clean speech in temporal domain, which is supposed to retain the complete speech information (including phase). Two exemplary works are shown in~\cite{Qian17-Speech} and~\cite{Rethage17-Wavenet}. Particularly in~\cite{Qian17-Speech}, an explicit prior model that learns the conditional distribution of speech samples for clean speech is further incorporated with WaveNet, to regularise the enhanced speech to be more speech-like. 

To further refine the model enhancement performance, adversarial training has recently attracted increasing attention. This training algorithm implements two networks, \ie one generative network (G) and one discriminative network (D), in a cascaded network structure. The generative network tries to map the noisy speech into the clean speech so as to fool the discriminative network, whereas the discriminative network aims to distinguish whether its inputs come from the enhanced speech (False) or the clean speech (True). Therefore, the two networks play a minimax game, and are optimised by 
\begin{equation}
\begin{aligned} 
\min_G \max_D V(D,G) = & \mathbb{E}_{\mb{x}\sim p_{data}(\mb{x})}[\text{log}(D(\mb{x}))] + \\
&\mathbb{E}_{\hat{\mb{x}}\sim p_{data}(\hat{\mb{x}})}[\text{log}(1-D(G(\mb{y})))].
\end{aligned}
\end{equation}
The adversarial training strategy was examined in~\cite{Pascual17-Speech} and~\cite{Michelsanti17-Conditional}, and was found to perform superior to other traditional approaches, such as Wiener filtering.  

\subsubsection{Brief Discussion}
The above reviewed works reflect a trend that the employed representations for enhancement have gradually moved from cepstral domain into temporal domain, mainly thanks to i) the powerful capability of deep learning to automatically extracted effective representations from raw data which ideally retain the complete information compared with the manually extracted features like MFCCs; ii) the advance of novel architecture of neural networks (\eg dilated CNN~\cite{Oord16-Wavenet}) which dramatically reduce the computational load; iii) the development of cloud computing which makes it possible to handle such a task. 

They also reflect another trend relating to the network training strategy, which starts to shift from traditional way with a single network to an adversarial way with two networks~\cite{Goodfellow14-Generative}. 
The adversarial training strategy regards the enhancement process as an image generation process, with the aid of a discriminative network to enhance the generative quality of the generative network. With recent rapid development of GAN in machine learning~\cite{Creswell17-Generative}, it can be expected that further improvements will be achieved in speech/feature enhancement in the future. 

However, while many works simply regard the spectrogram as a traditional visual image, few works specifically take their differences into account.  Traditional visual images are locally correlated, \ie nearby pixels are likely to have similar intensities and colours; whereas the spectrograms often include harmonic correlations which spread along frequency axis while local correlation may be weaker. Therefore, more efforts are required towards this direction.

\subsection{Masking-based Deep Enhancement Methods}
\label{subsubsec:masking-based}

Different from the mapping-based methods, masking-based methods aim to learn a regression function from a noisy speech spectrum $Y(n,f)$ to a Time-Frequency (T-F) mask $M(n,f)$. That is, 
\begin{equation}
Y(n,f)\xrightarrow{F} M(n,f).  
\end{equation}

\subsubsection{Masks}
Two most commonly used masks in the literature include: {\em binary}-based mask~\cite{Wang05-Ideal} and {\em ratio}-based mask~\cite{Srinivasan06-Binary}.
Typical binary-based mask often refers to {\em Ideal Binary Mask} (IBM), where a T-F mask unit is set to 1 if the local SNR is greater than a threshold $R$ (indicating clean speech domination), or 0 if otherwise (indicating noise domination). 
That is, 
\begin{equation}\label{eq:ibm}
M^b(n,f) = \left\{
             \begin{array}{lcl}
             1,\quad \text{if}\ \ SNR(n,f)>R, \\
             0,\quad \text{otherwise},
             \end{array}  
        \right.
\end {equation}
where $SNR(n,f)$ denotes the local SNR within the T-F unit at the frame index $n$ and the frequency bin $f$. Hence, the IBM is a binary matrix. 
Typical ratio-based mask often indicates the so-called {\em Ideal Ratio Mask} (IRM), where a T-F mask unit is assigned by the soft ratio of the clean speech and the noisy (mixture) speech, as follows: 
\begin{equation}\label{eq:irm}
M^r(n,f) = \frac{S^\alpha(n,f)}{S^\alpha(n,f)+N^\alpha(n,f)},
\end {equation}
where $S(n,f)$ and $N(n,f)$ are the magnitudes of clean speech and noise in the T-F domain, respectively, and $\alpha$ is a warping factor of the magnitudes in order to differentially affect the sharpness of the mask or the dynamic ranges of the features. Specifically, for example, if $\alpha$=2/3, 1, or 2, the IRM is calculated from an `auditory', magnitude, or power spectrum, respectively. 
When $\alpha=2$, the IRM is closely related to the Wiener filter, and can be viewed as its instantaneous version. 
From Eq.~(\ref{eq:ibm}) and (\ref{eq:irm}), it can be seen that IRM-based approaches could deliver a less distorted enhanced speech, while it could potentially involves much interference~\cite{Grais16-Combining}.

Wang and Wang~\cite{Wang13-Towards} first introduced DNNs to perform IBM estimation for speech separation, and reported large performance improvement over non-DNN-based methods. 
Subsequently, Wang \et~\cite{Wang14-training} compared a variety of masks and indicated that ratio masking (\eg IRM) is superior to binary masking (\eg IBM) in terms of objective intelligibility and quality metrics. 
This conclusion was further supported by the work in~\cite{Narayanan13-Ideal}, where the obtained results suggested that IRM achieves better ASR performance than IBM. Further, 
motivated by the advantages and disadvantages of IBM and IRM, Grais~\et~\cite{Grais16-Combining} combined the IBM- and the IRM-based enhanced (separated) speech by another neural network, to exploit the compensation between two approaches.  

Rather than estimating the masks in the T-F domain, the masking-based approaches were also successfully applied to a reduced feature space -- Mel-frequency domain~\cite{Narayanan13-Ideal,Weninger14-Discriminatively} and log-Mel-frequency domain~\cite{Weninger14-Single} that have frequently been proven to be effective for ASR in deep learning. The experimental results in~\cite{Weninger14-Discriminatively} showed that the masking-based approaches in the Mel-frequency domain perform better than the ones in the T-F domain in terms of SDR. 

Further, another trend in masking-based approaches is replacing DNNs with LSTM-RNNs as the mask learning model~\cite{Weninger14-Single,Weninger14-Discriminatively, Weninger15-Speech}, since LSTM-RNNs have shown to be capable of learning the speech and noise context information in a long temporal range 
thus also often being able to model events that appear non-stationary in the short term. The research efforts~\cite{Weninger14-Discriminatively} have demonstrated that LSTM-RNNs can notably outperform DBM/SAE alternatives in the mask estimation for source separation.

However, both the IBM and IRM-based approaches for calculating the target masks simply ignore the distorted phase information, even though it has been shown to be helpful for speech enhancement~\cite{Paliwal11-importance}.
For this reason, Erdogan~\et~\cite{Erdogan15-Phase} proposed a {\em Phase-Sensitive Mask} (PSM) that is calculated by 
\begin{equation}
 M^p(n,f) = \frac{|S(n,f)|}{|Y(n,f)|}cos(\theta), 
\end{equation}
where $\theta$ is the difference between the clean speech phase $\theta^s$ and the noisy speech phase $\theta^n$, \ie $\theta=\theta^s-\theta^n$. The experimental results on CHiME-2 database show that it outperforms the phase-nonsensitive approaches.

Note that, PSM does not completely enhance reverberant speech, since it cannot
completely restore the phase. For this reason, Williamson and Wang~\cite{Williamson17-Time} further developed this approach, naming it {\em complex IRM} (cIRM). It is defined as 
\begin{equation}
 M^c(n,f) = \frac{|S(n,f)|}{|Y(n,f)|}e^{j(\theta^s-\theta^y)}. 
\end{equation}
Therefore, cIRM can be regarded as the IRM in the complex domain, while PSM corresponds to the real component of the cIRM. Both two phase-based masks were demonstrated to be more effective than normal IRMs in suppressing the reverberated noise in~\cite{Williamson17-Time}.

\subsubsection{Objective Functions and Training Strategies}
In the neural network training stage, given the input $\mb{y}$ from the T-F domain of mixed noisy signals $Y(n,f)$ and the target $\mb{x}$ from the calculated T-F mask $M(n,f)$, the parameters of neural networks $\theta$ are determined by the so called {\em Mask Approximation} (MA) objective function. That is, it attempts to minimise the MSE between the estimated mask and the target mask as follows
\begin{equation}\label{eq:mse}
  \mathcal{J(\theta)}=\frac{1}{N}\sum_{n=1}^{N}\lVert F(\mb{y}_{n})-M(n,f)\rVert^2,
\end{equation}
where $\lVert\cdot\rVert^2$ is the squared loss, $n$ denotes the frame index, and $F(\mb{y}_{n})$ is restricted to the range [0,1].  

In the test stage, to filter out the noise, the estimated mask $\hat{M}(n,f)=F(\mb{y}_{n})$ is sequentially applied to the spectrum of the mixed noisy signal $\mb{y}$  by 
\begin{equation}
 \hat{\mb{x}}_n=\mb{y}_n\otimes \hat{M}(n,f), 
\end{equation}
where $\otimes$ denotes the elementwise multiplication. After that, it transforms the estimated clean spectrum $\hat{\mb{x}}$ back to the time-domain signal $\hat{s}(t)$ by an inverse STFT. 

Apart from the MA-based objective function, more and more studies have recently started to use {\em Signal Approximation} (SA) objective functions~\cite{Huang14-Deep,Weninger14-Discriminatively, Huang15-Joint}. Such an alternative straightforwardly targets minimising the MSE between the estimated clean spectrum $\hat{\mb{x}}=\mb{y}\otimes \hat{M}(n,f)$ and the target clean spectrum $\mb{x}$ by 
\begin{equation}\label{eq:mse}
  \mathcal{J(\theta)}=\frac{1}{N}\sum_{n=1}^{N}\lVert \mb{y}_n\otimes \hat{M}(n,f)-\mb{x}_n\rVert^2. 
\end{equation}
This is indeed similar to the objective function used for the mapping-based methods (cf.~Section~\ref{subsubsec:mapping-based}). Employing the SA based objective function was empirically examined to perform better than the MA-based one for source separation~\cite{Weninger14-Discriminatively}. Furthermore, the conclusions found in~\cite{Weninger14-Discriminatively} and~\cite{Wang15-deep} indicate that combining the two objective functions (\ie MA and SA) can further improve the speech enhancement performance in both the magnitude and the Mel-spectral domains. 

Due to the importance of phase information as aforementioned, Weninger \et~\cite{Weninger15-Speech} took the phase information in the objective function, which is called {\em Phase-sensitive SA} (PSA). Specifically, the network does not predict the phase, but still predicts a masking. However,  in the objective function (cf.~Eq.~(\ref{eq:mse})), the terms of $\mb{y}_n$ and $\mb{x}_n$ are in the complex domain, making the network learn to shrink the mask estimates when the noise is high~\cite{Weninger15-Speech}.

Additionally, a multi-task learning framework was proposed in~\cite{Huang14-Deep,Huang15-Joint} to jointly learn multiple sources (\ie speech and noise) and the mask simultaneously. The assumption behind this idea is that the relationship between noise and its caused speech distortion could be learnt and help for estimating the clean speech. The experimental results have shown that such a joint training framework is superior to the isolated training way~\cite{Huang14-Deep}. 

Although the masking-based approaches were initially designed for removing additive noise, recent research has showed that they are capable of eliminating convolutional noise as well~\cite{Weninger14-Single,Erdogan15-Phase,Weninger15-Speech}.

\begin{table*}[t]
\centering
\caption{A summary of representative {\em single-channel} approaches based on deep learning for environmentally robust speech recognition.  Those methods are summarised at different ASR processing stages ({\em front}-end, {\em back}-end, and {\em joint} front- and back-end).}
\label{tab:single-channel}
\resizebox{\textwidth}{!}{  
\begin{threeparttable}
\begin{tabular}
{p{0.04\textwidth}p{0.1\textwidth}p{0.13\textwidth}p{0.4\textwidth}p{0.4\textwidth}}
\toprule
stage & approaches  & typical publications & \qquad advantages   & \qquad disadvantages    \\

\midrule
\multirow{5}{*}{\rotatebox[origin=c]{90}{front (mapping)}} 
&MFCC&\cite{Maas12-Recurrent,Woellmer13-Feature,Weninger13-Munich,Zhang14-Channel,Feng14-Speech} &low dimension, require less computational load&lose much information, (almost) irreversible to raw signals  \\ 
&(log) Mel&\cite{Lu13-Speech,Ishii13-Reverberant,Weninger14-Feature,Weninger14-MERL}&low dimension, require less computational load & lose some information, (almost) irreversible to raw signals \\ 
&(log/power) mag.&\cite{Xu14-NMF,Xu15-Regression,Han15-Learning,Park16-Fully}&invertible to the audio signal& high dimension, require high computational load, each element is equally important \\
&temporal&\cite{Qian17-Speech,Rethage17-Wavenet}&retain the complete information&large data size, require heavy computational load \\ 

\midrule
\multirow{8}{*}{\rotatebox[origin=c]{90}{front (masking)}} 
&IBM&\cite{Wang13-Towards,Wang14-training,Grais16-Combining}&little interference&much magnitude distortion \\
&IRM&\cite{Narayanan13-Ideal,Weninger14-Discriminatively,Wang14-training,Huang15-Joint,Grais16-Combining}&little magnitude distortion&much interference\\
&PSM&\cite{Erdogan15-Phase}&less phase distortion and little interference &do not restore the complete phase information\\
&cIRM&\cite{Williamson17-Speech}&learn a complete relationship between noisy and clean speech for both magnitude and phase&relative complex to compute\\
\cdashline{2-5}
&MA&\cite{Narayanan13-Ideal,Wang13-Towards,Weninger14-Discriminatively}&most straightforward way&no enhanced phase\\
&SA&\cite{Huang15-Joint,Grais16-Combining,Erdogan15-Phase}&directly optimise the objective& no enhanced phase\\
&PSA&\cite{Weninger15-Speech}&considered phase information when  predicting mask& \\

\midrule
\multirow{15}{*}{\rotatebox[origin=c]{90}{back}}  
&tandem &\cite{Sharma00-Feature,Wollmer09-Robust}&\multirow{3}{*}{$\Bigg\}$\parbox{0.35\textwidth}{explicitly make use of discriminative features}} &\multirow{3}{*}{\parbox{0.4\textwidth}{$\Bigg\}$HMM model dependent}} \\
&double-stream &\cite{Weninger14-MERL,Geiger14-Memory}&& \\ 
&hybrid &\cite{Geiger14-Robust}&& \\ 
&multi-condit. train. &\cite{Wang13-Towards,Seltzer13-investigation}&straightforward and very efficient&require many data in different noisy scenarios \\ 
&model adapt. &\cite{Mirsamadi15-study}&flexible to different noisy environments&require a relatively large amount of adaptation data, otherwise easy overfitted \\ 
&NAT &\cite{Seltzer13-investigation,Karanasou14-Adaptation,Yu15-Robust}&easy to be implemented& require another disassociated model to estimate noise and cannot be optimised jointly \\ 
&dynamic NAT &\cite{Xu14-Dynamic}&more efficient to deal with non-stationary noise&more efforts to estimate noise \\ 
&multi-task train. &\cite{Giri15-Improving,Chen15-Speech}&exploit the clean-sensitive speech &could lose some discriminative features \\ 

\midrule
\multirow{10}{*}{\rotatebox[origin=c]{90}{joint}}   
&re-training&\cite{Weninger14-Feature,Weninger13-Munich}&no need to change the structure of acoustic model&do not guarantee a better speech recogniser since the two nets are optimised by different metrics \\ 
&joint&\cite{Lee16-Two,Lee17-Integrated,Gao15-Joint,Wang16-Joint,Mimura16-Joint,Ravanelli17-Network}&exploit the complementary of enhancement networks and speech recognition networks&tricky to combine the two networks \\ 
&end-to-end&\cite{Amodei16-Deep,Qian16-Very}&automatically distil the salient features for speech recognition from raw noisy speech (or low-level features), so it reduces the information loss&require a large amount of training data and heavy computational load \\ 
\bottomrule
\end{tabular}
\end{threeparttable}}
\end{table*}

\section{Back-End Techniques}
\label{sec:back-end}

The back-end techniques are also known as {\em model-based} techniques. They leave the noisy observation unchanged, and instead let the neural networks automatically find out the relationship between the observed noisy speech and the phonetic targets. Compared with the aforementioned front-end techniques, a drawback of the back-end techniques is that they have to change the parameters, or even structures, of previous trained Acoustic Model (AM). 

Early works focus on the improvement of the model structure in order to make it more robust to recognise noisy speech. The most popular approaches involve with a combination of DNNs and HMMs, such that they take advantage of neural networks for discriminative classification and HMM for context learning. 
A \textit{tandem} structure is one typical approach proposed in~\cite{Sharma00-Feature}. It utilises the neural networks to predict phonemes, and explicitly considers the phoneme prediction as a discriminative feature and combines it with original features for HMM to make a final prediction. 

{\em Multi-stream} HMM architecture~\cite{Geiger14-Memory} is another popular approach to incorporate DNN with traditional GMM-HMM model. Specifically, given the HMM emission state $s$ and the input vector $\mb{y}$, at every time frame $n$, the double-stream HMM has access to two independent information sources, $p_G(\mb{y}_n|s_n)$ and $p_L(\mb{y}_n|s_n)$, the acoustic likelihoods of the GMM and the RNN predictions, respectively. In particular, the RNN-based AM is discriminatively trained to generate frame-wise phone predictions.  The double-stream emission probability is computed as 
\begin{equation}\label{eq:multistream}
 p(\mb{y}_n|s_n) = p_G(\mb{y}_n|s_n)^\lambda p_L(\mb{y}_n|s_n)^{1-\lambda}, 
\end{equation}
where the variable $\lambda\in[0,1]$ denotes the stream weight. 
This approach combines GMM and DNN to leverage the reliable adaptation performance of the former and high quality classification performance of the latter. 
In particular, when setting $\lambda$ to 0, the structure is known as {\em hybrid} NN/HMM model, where only the likelihoods of the NN predictions are employed for HMMs. 

Recently, owing to the  capability of LSTM-RNN in learning long-term dependence, the frequently used fully-connected layers in DNN have been shifted to LSTM layers in either tandem~\cite{Wollmer09-Robust}, double-stream~\cite{Geiger14-Memory}, or hybrid structures~\cite{Geiger14-Robust}.

One main drawback of the combined NN/HMM structures is that it highly depends on the HMM model that, nevertheless, is gradually losing its ground in speech recognition, and being replaced by the rapidly developed DNN model-only~\cite{Amodei16-Deep}. Therefore, HMM-independent approaches are more than necessary than ever before. 
The widely used approach comes to {\em multi-condition} training~\cite{Seltzer13-investigation}. In doing this, various acoustic variations caused by different noises are provided in the training process, reducing the acoustic distribution mismatch between the training and the test speech. However, it requires a large amount of data in various noisy conditions, which is rarely the case in practise. 

To release the large-data-size requirement and make the model become flexible, another common way is {\em model adaptation}, which aims to modify the parameters of a pre-trained AM to compensate the acoustic distribution mismatch. 
However, modifying the entire weights of the neural networks (AM) with small adaptation data easily leads to overfitting and results in noise-dependent parameter sets~\cite{Mirsamadi15-study}. Alternatively, a part of neural network parameters can be modified. 
For example, the authors of the work~\cite{Mirsamadi15-study} added an extra layer with linear activations to the input layer, the hidden layers, or the output layer of neural networks, for model adaptation, which contributes to a considerable system robustness in environmentally noisy conditions.  

Rather than forcing the pre-trained AM to adapt to various noisy conditions, an alternative approach aims to let the network-based AM be informed about the noise information (or acoustic space information~\cite{Giri15-Improving}) when training, which is often termed as {\em Noise-Aware Training} (NAT)~\cite{Seltzer13-investigation}. In this case, 
a noise estimation $\hat{\mb{n}}$ presented in the signal serves as an auxiliary input and is incorporated with the original observation input $\mb{y}$, \ie $[\mb{y},\hat{\mb{n}}]$. 
In this way, the DNN is being given additional cues in order to automatically learn the relationship between noisy speech and noise, which is beneficial to predict phonetic targets~\cite{Seltzer13-investigation}. Experimental results on the `Aurora-4' database show that the NAT-based AM is quite noise robust. 

Therefore, the key point is changed to how to represent the noise information. 
Early works implement traditional signal processing approaches, such as MMSE, and estimate the noise over each sentence. Recently, a more general way to represent noise is employing {\em i-vectors}~\cite{Dehak11-Front}, which were originally developed for speaker recognition. The i-vector can be calculated either from the hand-crafted features such as commonly used MFCCs~\cite{Karanasou14-Adaptation}, or from the automatically extracted bottleneck representations by DNN~\cite{Yu15-Robust}; and either from the raw noisy features, or from the enhanced features, \eg Vector Taylor Series (VTS)~\cite{Yu15-Robust}. 

Most of these studies assume that the noise is stationary within an utterance, so that the obtained noise estimation or i-vector can be applied to the whole utterance. Nevertheless, this is not always the case in practise. To address this issue, dynamic NAT was introduced in~\cite{Xu14-Dynamic}, where the authors used masking-based approaches (cf.~Section~\ref{subsubsec:masking-based}) to estimate the noise that varies along time. This approach performs more efficient especially for non-stationary noise, whereas it requires an extra DNN for noise estimation. 

Apart from these approaches, a {\em multi-task learning} based AM has attracted increasing attention. For example, the work done in~\cite{Giri15-Improving} and \cite{Chen15-Speech} respectively introduced similar multi-task learning architectures but different network types (\ie one is a feed-forward DNN and the other one is a LSTM-RNN)  
for noisy speech recognition, where the primary task is the senone classification and the augmented task is reconstructing the clean speech features. In these architectures, the objective function is calculated by  
\begin{equation}
 \mathcal{J}(\mb{\theta}) = \lambda E_c + (1-\lambda) E_r,
\end{equation}
where $E_c$ and $E_r$ indicate the senone classification error and the clean feature reconstruction error, respectively. The underlying assumption of this approach is that the representations that are good for producing clean speech should be easier to be classified.

\section{Joint Front- and Back-End Training Techniques} 
\label{sec:end-to-end}

Most research efforts on flighting with the environmental noise in the past few year were separately made on the system front end or back end. That is, speech/feature enhancement and speech recognition are often designed independently and, in many cases, the enhancement part is tuned according to the metrics such as segSNR, SDR, and PESQ, which are not directly correlated with the final recognition performance. 

To address this issue, a straightforward way is employing the enhanced speech obtained in the front end to {\em re-train} the pre-trained AM in the back-end~\cite{Weninger13-Munich}. This simply remains everything unchanged but a further re-training process on the AM. 

A more sophisticated {\em joint} DNN structure was proposed in~\cite{Lee16-Two,Lee17-Integrated}, where the authors concatenated two independent pre-trained DNNs. The first DNN performs the reconstruction of the clean features from noisy features augmented by a noise estimation. The second DNN attempts to learn the mapping between the reconstructed features and the phonetic targets~\cite{Lee16-Two}. Then, joint the two individual networks as one and further fine-tune the network parameters together.  
Compared with the re-training strategy, the joint neural networks could learn more discriminative representations for speech recognition when reconstructing the clean features from the noisy ones by feature enhancement in the front-end. 

Furthermore, the work done in~\cite{Narayanan14-Joint} even left out the pre-training process, and directly concatenated a DNN-based speech separation front-end and a DNN-based AM back-end to build a large neural network, and jointly adjusted the whole weights. 
In doing this, the enhancement front-end is able to provide enhanced speech desired by the acoustic model and the acoustic model can guide the enhancement front-end to produce more discriminative enhancement. In other words, the linguistic information contained in the acoustic models can flow back to influence the enhancement front-end at the training stage. Similar work was further done in~\cite{Gao15-Joint,Wang16-Joint,Mimura16-Joint}. 

Despite the considerable effectiveness of such joint training frameworks, the enhancement process and the speech recognition process suffer from a uni-directional communication. To this end, a novel architecture was proposed in~\cite{Ravanelli17-Network}. It jointly optimises the enhancement network and speech recognition network in a parallel way, rather than a cascaded way; the activations of the hidden layer of each network will be mutually concatenated as new inputs of their next hidden layer. Thus, all the components of two networks are jointly trained and better cooperate with each other. 

More recently, an \textit{end-to-end} architecture has attracted dramatic attention, and have shown great promise in latest ASR systems~\cite{Sainath15-Convolutional,Amodei16-Deep}. Its central idea is to jointly optimise the parameters of the networks at the front-end which automatically learn the inherent representations from low-level features/signals for the task at hand, and the networks at the back-end which provide final predictions. For noisy speech recognition, a quite recent and  well-developed framework has been reported in~\cite{Qian16-Very}, where two tasks were evaluated: the Aurora-4 task with multiple additive noise types and channel mismatch, and the `AMI' meeting transcription task with significant reverberation. In this framework, a variety of very deep CNNs with many convolutional layers were implemented, and each of them is followed by four fully-connected layers and one softmax output layer for senone prediction. Compared with DBMs, the CNNs have the advantages~\cite{Qian16-Very}: 1) they are well suited to model the local correlations in both time and frequency in speech spectrogram; and 2) translational invariance, such as the frequency shift due to speaker or speaking style variations, can be more easily captured by CNNs. The reported results on the AMI corpus by using the proposed end-to-end framework is much higher than the results of traditional DBMs and are competitive to the LSTM-RNN-based AM; and the results on Aurora-4 beat any other published results on this database, even without performing any speech and feature enhancement approaches.

\section{Multi-Channel Techniques}
\label{sec:multi-channel}

Microphone arrays and {\em multi-channel} processing techniques have recently played an increasingly significant role in the development of robust ASR~\cite{Barker15-third, Kinoshita16-summary}. 
A central approach is {\em acoustical beamforming}, \ie spatio-temporal filtering that operates on the outputs of microphone arrays and converts them to a single-channel signal while amplifying the speech from the desired direction and attenuating the noise coming from other directions. 
The beamformer output is often further enhanced by a \emph{microphone array post-filter} \cite{Marro98-Analysis, McCowan03-Microphone}. After that, the back-end techniques for single-channel speech can be applied to this enhanced data for speech recognition. 

With the rapid development, deep learning has emerged as a powerful tool to evolute the traditional methods. 
In the following, we separately discuss the latest deep learning approaches either in a \emph{supportive} way to assist traditional beamforming methods (a well known survey can be found in~\cite{Van88-Beamforming}) and post-filtering methods in the front end, or an \emph{independent} way to address the multi-channel speech recognition in a joint front and back end. Note that we do not summarise the back-end techniques in this section since it shares the same techniques with the ones for single channel as aforementioned.  

\begin{table*}[t]
\centering
\caption{A summary of representative {\em multi-channel} approaches based on deep learning for environmentally robust speech recognition. Those methods are summarised at different ASR processing stages ({\em front}-end and {\em joint} front- and back-end).}
\label{tab:multi-channel}
\resizebox{\textwidth}{!}{  
\begin{threeparttable}
\begin{tabular}
{p{0.04\textwidth}p{0.2\textwidth}p{0.1\textwidth}p{0.4\textwidth}p{0.4\textwidth}}
\toprule
stage & approaches  & typical publications & \qquad advantages   & \qquad disadvantages    \\
\midrule
\multirow{4}{*}{\rotatebox[origin=c]{90}{front}} 
&mask estimation &\cite{Heymann15-BLSTM,Heymann16-Neural,Erdogan16-Improved,Menne16-RWTH} & avoid relying on a DOA estimation&require large-scale training data \\ 
&filter coefficients estimation &\cite{Xiao16-Deep,Li16-Neural,Meng17-Deep} &easily to be integrated with DNN AM as a joint network & based on the simulated data in all possible scenarios \\ 
&post-filter estimation &\cite{Pertilae14-Microphone} & do not require explicit estimates of the signal and noise statistics & require large-scale simulated data \\ 

\midrule
\multirow{8}{*}{\rotatebox[origin=c]{90}{joint}} 
&channel concatenation &\cite{Swietojanski13-Hybrid,Liu14-Using} &require no knowledge of microphone array geometry and signal information & unclear on a severe mismatch among multiple channels\\ 
&cross-channel max-pooling &\cite{Swietojanski14-Convolutional} &able to pick the most informative channel & unable to make use of the spatial information found in multi-channel signals\\ 
&factoring spatial \& spectral filtering &\cite{Sainath17-Multichannel} &robust to varying target speaker direction of arrival & additional computational cost \\ 
&end-to-end &\cite{Hoshen15-Speech,Ochiai17-Multichannel} & automatically extracted the underlying and salient representations over multiple channels& heavy parameters tuning and computational load \\ 
\bottomrule
\end{tabular}
\end{threeparttable}}
\end{table*}

\subsection{Front End: NN-Supported Beamformers and Post-Filters}
Beamformers in general require a Direction-Of-Arrival (DOA) estimate for the target signal. In {\em Delay-and-Sum (DS) beamforming}, which is one of the simplest approaches and applies a fixed delay operation to align the signals of the different microphones before summing them, so as to focus on the desired target direction. 
In contrast, \emph{adaptive} beamformers update the filter coefficients based on estimates of the noise and signal statistics, and have now become the dominate approaches to address the non-stationary noise due to its time-varying attribute. Among them, the {\em Minimum Variance Distortionless Response (MVDR)} approach and the {\em Generalised EigenValue (GEV)} approach have shown to be particularly promising recently~\cite{Barker15-third,Vincent16-analysis}. 

Specifically, the MVDR beamforming works in the frequency domain and aims to minimise the energy at the beamformer output, while simultaneously keeping the gain in the direction of the target signal fixed at unity. 
The complex-valued signal model is $\mb{Y}(n) = S(n)\mb{d} + \mb{A}(n)$, where the vector $\mb{Y}(n) = (Y_{1}(n),\ldots,Y_{M}(n))^{T}$ contains the instantaneous noisy observations at the $n$-th time instant on a given discrete frequency bin as registered by the $M$ microphones, $S(n)$ is the corresponding complex frequency bin of the unknown transmitted signal, the steering vector $\mb{d}$ is the desired signal spatial signature encoding its direction of arrival and $\mb{A}(n)$ is a $(M \times 1)$ vector containing the noise and interference contributions. Both the signal and the noise are assumed to have zero mean. In operation, the beamformer computes a linear combination of a complex weight vector $\mb{w}$ and the observation vector $\mb{Y}(n)$ as
$\mb{x}(n) = \mb{w}^{H}\mb{Y}(n)$, 
where $(\cdot)^H$ denotes the Hermitian transpose. 
In determining $\mb{w}$ using the MVDR criterion, the spatial covariance matrix representing the covariance of the noise plus interference will be needed. It is generally unknown but can be estimated as a sample covariance matrix of a suitable segment of $N$ observations as $\mb{R}_{VV} = (1/N) \sum_{n}\mb{Y}(n)\mb{Y}^{H}(n)$ \cite{Mestre03-diagonal}. By then minimising $\mb{w}^{H}\mb{R}_{VV}\mb{w}$ with respect to $\mb{w}$ subject to the constraint $\mb{w}^{H}\mb{d}=1$, as mentioned above, the MVDR beamformer filter coefficients are given by \cite{Cox87-Robust}

\begin{equation} \label{eq:mvdr-beamformer-solution}
\hat{\mb{w}}_{MVDR} = \frac{\mb{R}_{VV}^{-1}\mb{d}}{\mb{d}^{H}\mb{R}_{VV}^{-1}\mb{d}}.
\end{equation}

The MVDR beamformer is not robust against an inaccurately estimated steering vector $\mb{d}$ \cite{Khabbazibasmenj12-Robust}. In contrast, GEV beamformer requires no DOA estimate and is based on maximising the output signal-to-noise ratio \cite{Warsitz07-Blind}. The beamformer filter coefficients for a given frequency bin are found as the principal eigenvector of a generalised eigenvalue problem as required by~\cite{Warsitz07-Blind}

\begin{equation} \label{eq:gev-beamformer-criterion}
 \hat{\mb{w}}_{GEV} = \argmax_{\mb{w}}\frac{\mb{w}^H\mb{R}_{SS}\mb{w}}{\mb{w}^H\mb{R}_{VV}\mb{w}},
\end{equation}
where $\mb{R}_{SS}$ and $\mb{R}_{VV}$ are the required estimates of the spatial covariance matrices of the target speech and noise/interference, respectively.

Recently reported NN-supported beamformers can be generally categorised into two types~\cite{Ochiai17-Multichannel}: (i) beamformers with a \emph{mask estimation} network \cite{Heymann15-BLSTM,Heymann16-Neural,Heymann16-Wide,Menne16-RWTH}; and (ii) beamformers with a \emph{filter estimation} network \cite{Xiao16-Deep,Li16-Neural,Meng17-Deep}. Both approaches aim to obtain an enhanced signal based on the formalisation of the conventional filter-and-sum beamformer in the time-frequency domain. 
The difference between them is how the filter coefficients are generated by neural networks. The former approach uses neural networks to estimate noise or speech masks (cf.~Section~\ref{subsubsec:masking-based}), which are then applied to calculate the spatial covariance matrix further followed by a calculation of filter coefficients by Eq.~(\ref{eq:mvdr-beamformer-solution}) and (\ref{eq:gev-beamformer-criterion}). On contrary, the later approach skips a series of interval process. It directly utilises neural networks to estimate the filter coefficients. In both approaches, the estimated filter coefficients are then applied to the multi-channel noisy signals to obtain the enhanced speech signals. 

Specifically, the mask estimation-based beamformer was firstly investigated in~\cite{Heymann15-BLSTM}, where LSTM-RNNs were used to 
estimate two IBMs for each microphone channel: The two IBMs receptively indicate for each T-F bins whether they are presumably dominated by speech or noise. To train neural networks, the authors further used a multi-task learning framework; the inputs are noisy speech, and the targets are two IBMs. These obtained masks are then condensed to a single speech and a single noise mask by a median filter, which are sequentially used for estimating the spatial covariance matrices $\mb{R}_{SS}$ and $\mb{R}_{VV}$ and in turn the beamformer coefficients $\hat{\mb{w}}_{GEV}$. 
However, this approach requires {\em both} speech and noise counterparts of the noisy speech for each microphone channel. In this case, only simulated data is possible to be employed for the network training. To relax this requirement to some extent, a follow-up work has been presented in~\cite{Heymann16-Neural}, where only the clean speech was employed for the mask estimation. This slight improvement enables one to utilise more realistic noisy and clean speech pairs, which can be recorded simultaneously by a close microphone (for clean speech) and a distant microphone array (for noisy speech). The experimental results shown in~\cite{Heymann16-Neural} were competitive with the ones in previous work~\cite{Heymann15-BLSTM}. Apart from the mask estimation for GEV, similar approach was also applied to MVDR, where the steering vector is calculated by the principal component of the estimated spatial covariance matrix of speech, \ie $\mb{d}=\mathcal{P}(\mb{R}_{VV})$.
The effectiveness of all these mentioned approaches has been demonstrated in the fourth CHiME Challenge~\cite{Menne16-RWTH,Heymann16-Wide}.

A typical filter coefficients-based approach was evaluated in~\cite{Xiao16-Deep}, where the networks were trained with generalised cross correlation from simulated multi-channel data from a given array geometry using all possible DOA angles. As conventional neural networks are not able to handle complex values directly, the real and imaginary parts of each complex weight are predicted independently~\cite{Xiao16-Deep}. Similar investigation was also shown in~\cite{Meng17-Deep,Li16-Neural}. 

As for post-filtering, very few recent papers appear to have used neural networks for this purpose. One such study evaluated a non-deep MLP network in predicting the post-filter parameters for a circular microphone array \cite{Pertilae14-Microphone}.

\subsection{Joint Front- and Back-End Multi-Channel Techniques}

Rather than using neural networks to support traditional beamformers and post-filters for speech enhancement, joint front- and back-end multi-channel ASR systems have recently attracted considerable attention with a goal of decreasing the WER directly~\cite{Liu14-Using,Swietojanski14-Convolutional,Hoshen15-Speech}. 
In~\cite{Swietojanski13-Hybrid}, the individual features extracted from each microphone channel are concatenated as a long single feature vector and fed into a DNN for AM. Whilst such a feature concatenation operation is simple, it was still found to be effective for dereverberation on the AMI dataset~\cite{Swietojanski13-Hybrid}, and was further verified in~\cite{Liu14-Using}.  

A more sophisticated approach was proposed in~\cite{Swietojanski14-Convolutional}. In this work, the authors utilised a joint network structure of several individual convolutional layers followed by a shared fully-connected feedforward network. In more detail, each individual convolutional layer was operated on each channel independently with the magnitude spectrum as input, and a max pooling was proceeded across channels to choose the channel with the largest response in each node. This algorithm performs better than the one by applying a  CNN after a DS beamformer~\cite{Swietojanski14-Convolutional}.

\begin{table*}[!th]
\centering
\caption{Benchmarks for four selected standard corpora (\ie Aurora-4, CHiME-2, CHiME-4, and AMI). Note that only the deep learning related approaches were indicated for each present system. That is, many other traditional approaches might also be utilised. Further note that DNN mentioned in the table generally refers to Deep Boltzmann Machine (DBM) or Deep Belief Network (DBN). MCT: multi-condition training; MTL: multi-task learning; WRN: wide residual network~\cite{Zagoruyko16-Wide}; VDCNN: very deep CNN. }
\resizebox{\textwidth}{!}{  
\begin{threeparttable}
\begin{tabular}
{p{0.05\textwidth}p{0.2\textwidth}p{0.2\textwidth}p{0.1\textwidth}p{0.02\textwidth}p{0.15\textwidth}p{0.15\textwidth}p{0.07\textwidth}p{0.1\textwidth}}
\toprule
pub.             & \multicolumn{3}{c}{single channel}      && \multicolumn{2}{c}{multiple channels}  & model       & WER      \\
\cline{2-4} \cline{6-7}
    & front-end           & back-end        & joint      && front     & joint        &      & (or SDR)   \\

\midrule
\multicolumn{9}{c}{\bf{Aurora-4}} \\ 
\cite{Narayanan13-Ideal}           & IBM/IRM (MA; Mel)           & MCT                                &            && &                           & DNN                   & 16.50\%                    \\
\cite{Seltzer13-investigation}     &                             & NAT; MCT; re-training              &            && &                           & DNN                   & 12.40\%                    \\
\cite{Kundu16-Joint}               &                             & NAT, MTL                           &            && &                           &                       & 8.80\%                     \\
\cite{Qian16-Very}                 &                             &                                    & end-to-end && &                           & VDCNN         	 & 8.81\%                     \\
[10pt]

\multicolumn{9}{c}{\bf{CHiME-2}} \\ 
\cite{Weninger13-Munich}            & mapping: MFCC              & Re-training                                             &            &&  &                           & BLSTM                 & 26.73\%                    \\
\cite{Geiger14-TUM}                 &                            & multi-stream,re-training                                &            &&  &                           & BLSTM                 & 41.42\%                    \\
\cite{Weninger14-Discriminatively}  & masking: IRM (MA, SA; Mel) &                                                         &            &&  &                           & LSTM                  & 17.68 (SDR)                \\
\cite{Geiger14-Robust}              &                            & hybrid                                                  &            &&  &                           & BLSTM                 & 22.20\%                    \\
\cite{Weninger14-Feature}           & mapping: log Mel           & re-training                                             &            &&  &                           & BLSTM                 & 22.16\%                    \\
\cite{Erdogan15-Phase}              & masking: PSM (SA; log Mel) &                                                         &            &&  &                           & BLSTM                 & 14.76 (SDR)                \\
\cite{Han15-Learning}               & mapping: log mag           & hybrid                                                  &            &&  &                           & DNN                   & $\approx$25\%                 \\
\cite{Chen15-Speech}                & masking: IRM (SA; log Mel) & MTL; hybrid                                             &            &&  &                           & BLSTM                 & 16.04\%                    \\
\cite{Narayanan15-Improving}        & masking: IRM (MA; Mel)     &                                                         & joint      &&  &                           & DNN                   & 15.40\%                    \\
\cite{Weninger15-Speech}            & masking: IRM (PSA; Mel)    &                                                         &            &&  &                           & BLSTM                 & 13.76\%                    \\
\cite{Wang16-Joint}                 & masking: IRM (power spec.) & multi-stream; model adapt.; MCT			   & joint      &&  &                           & DNN                   & 10.63\%                    \\
[20pt]

\multicolumn{9}{c}{\bf{CHiME-4}\tnote{a}} \\ 
\cite{Menne16-RWTH}                 &&                           &            && mask est.                                 &                           & BLSTM                 & //; 4.0\%,5.2\%            \\
\cite{Heymann16-Wide}               &&                           &            && mask est.                                 &                           & WRN \& BLSTM          & 1.7\%,9.9\%; 3.1\%,3.9\%   \\
\cite{Xiao16-Study}                 &&                           &            && filter coeff. est.; mask est. 	           &                           & LSTM                  & 21.4\%,20.9\%; 5.0\%,6.4\% \\
\cite{Erdogan16-Multi}              && re-training; hybrid       &            && mask est.                                 &                           & BLSTM                 & //; 3.4\%,4.4\%            \\
\cite{Qian16-SJTU}                  && NAT                       &            &&                                           & end-to-end                & VDCNN \& LSTM 	       & 12.9\%,13.9\%; 6.3\%,6.4\% \\
[10pt]

\multicolumn{9}{c}{\bf{AMI}\tnote{b}}\\
\cite{Swietojanski13-Hybrid}        && MCT                      &            &&                                            & channel concatenation     & DNN-HMM               & 57.30\%   \\
\cite{Swietojanski14-Convolutional} &&                          &            &&                                            & cross-channel max-pooling & CNN                   & 49.40\%   \\
\cite{Liu14-Using}                  &&                          &            &&                                            & channel concatenation     & DNN                   & 44.80\%   \\
\cite{Qian16-Very}                 &&                          & end-to-end &&                                            &                           & VDCNN         	       & 46.90\%   \\
\cite{Xiao16-Deep}                  &&                          &            && filter coeff. est.                         & end-to-end                & DNN                   & 42.20\%   \\
\cite{Ochiai17-Multichannel}        &&                          &            && mask est.                                  & end-to-end                &                       & 39.00\%   \\
\bottomrule

\\
\multicolumn{9}{l}{[a] results are provided for the one and six channel signals (separated by `;') on the simmulated and real subsets of the evaluation dataset (separated by `,'); } \\
\multicolumn{9}{l}{[b] only the results for MDM subset are provided.}

\end{tabular}
\end{threeparttable}}
\label{tab:benchmark}
\end{table*}

Encouraged by this work as well as the research trend of end-to-end ASR systems, this work was extended to handle raw speech directly and without the operation of cross-layer max pooling~\cite{Hoshen15-Speech}. The advantage of these extensions is that the system can automatically exploit the spatial information found in the fine time structure, which primarily lies in the previously discarded FFT phase value, of the multichannel signals~\cite{Hoshen15-Speech}. 
A follow-up work was reported in~\cite{Sainath17-Multichannel}, where the authors employed two convolutional layers, instead of one layer, at the front-end. The assumption is that the spatial and spectral filtering operations can be separately processed by two convolutional layers. That is, the first layer is designed to be spatially selective, and the second layer is implemented to decompose frequencies that are shared across all spatial filters. By factoring the spatial and the spectral filters as separate layers in the network, the performance of the investigated system was notably improved in terms of WER~\cite{Sainath17-Multichannel}.

\section{Conclusions} 
\label{sec:conclusion}

In this survey, we have attempted to provide a comprehensive overview on the state-of-the-art and most promising deep learning approaches with the goal of improving the environmental robustness of speech recognition systems. These technologies are mainly introduced from the viewpoint of single-channel and multi-channel processing at different stages of the ASR processing chain, \ie the front-end, the back-end, or the joint front- and back-end. 

To intuitively compare the performance of different approaches, we selected four benchmark databases from Table~\ref{tab:database}, \ie Aurora, CHiME-2, CHiME-4, and AMI. The rationale behind this choice is that  Aurora-4 includes a large vocabulary among the Aurora series corpora; CHiME-2 is more frequently used in the past few years in comparison with CHiME-1 in single and two channels; CHiME-4 is the most recently employed standard database compared with CHiME-3 to address both additive and convolutional noise in multiple channels; and AMI has large-scale data compared with all aforementioned databases and is public available compared with Voice Search. The experimental results for each benchmark database are shown in Table~\ref{tab:benchmark}. 


From the table, we can find that i) the deep learning-based robust ASR systems are shifting from taking conventional hand-crafted features (\eg MFCCs) as the input, to automatically extracting the representative and discriminative features directly from noisy raw speech, mainly due to the fact that raw speech signals keep the entire information (\eg phase) related to the targets (\ie phoneme or word); ii) separate front-end and back-end systems are gradually defeated by joint, even end-to-end training systems, owing to the powerful non-linear learning capability of deep neural networks which can optimise all processing stages simultaneously; iii) the importance of multi-channel approaches is more striking considering the promising performance they offer. 

Due to the growth in popularity of microphones embedded in smartphones for example, more realistic and large size of data are increasingly utilised to train speech recognition model for flighting with the diverse and severe acoustic environments.  
This consequently requires more complex and deep neural network structures and high computing resources. 

Despite great achievements that deep learning has accomplished in the fast few years, 
as shown in the literature, an obvious performance gap still remains between the state-of-the-art noise-robust system and the one evaluated in a degradation-free, clean environment. Therefore, further efforts are still required for speech recognition to overcome the adverse effect of environmental noises~\cite{Barker15-third,Kinoshita16-summary,Vincent16-analysis}. 
We hope that this review could help researchers and developers to stand on the frontier of the developments in this field and to make greater breakthroughs.

\section*{Acknowledgements}
\noindent

This work was supported by Huawei Technologies Co.~Ltd.

\bibliographystyle{IEEEtran}
\bibliography{deepEnhancement}%

\end{document}

%% file: ASRFramework.tex
%
%
%
%
%

\tikzstyle{object} = [rectangle, draw,  text width=1.6cm, rounded corners, font=\small]
\tikzstyle{action} = [draw, ellipse,  text width=1.5cm,font=\small]
\tikzstyle{symbol} = [text centered, text width=2cm,]

\tikzstyle{arrow0} = [color=gray,->, line width=0.8pt]
\tikzstyle{arrow1} = [color=red,->, line width=0.8pt]
\tikzstyle{arrow2} = [color=blue,->, line width=0.8pt]

\tikzset{every node/.style={inner sep=-3pt,minimum height=0.9cm, text centered}}

\tikzstyle{vecArrow} = [line width=1.4pt,gray, decoration={markings, mark=at position
   1 with {\arrow[line width=1.4pt, gray]{open triangle 60}}},
   double distance=3pt, shorten >= 7pt,
   preaction = {decorate}]

\makebox[\linewidth]{%
\begin{tikzpicture}[>=stealth,node distance = 2.3cm,auto]
    \node [object] (prepro) {pre-processing};
    \node [object, right of=prepro,node distance = 1.8cm] (fe) {feature extraction};
    \node [action, right of=fe] (recogniser) {recogniser}; 
    \node [object, above left=0.7cm and -.6cm of recogniser,text width=1.5cm] (am) {acoustic model};
    \node [object, above right=0.7cm and -.6cm of recogniser,text width=1.5cm] (lm) {language model};
    \node [symbol, right of=recogniser,node distance = 1.6cm,text width=1cm] (text) {text}; 
    \node [symbol, left of=prepro, node distance = 1.8cm] (speech) {speech\\signal};
    
    \foreach \name / \y in {-3mm,-1.5mm,0mm,1.5mm,3mm}
      \draw[arrow0] ($(prepro.west)+(-5mm,\y)$) -- ($(prepro.west)+(0,\y)$); 
    
    \draw[arrow0] (prepro) -- (fe);
    \draw[arrow0] (fe) -- (recogniser);
    \draw[arrow0] (recogniser) -- (text);
    \draw[arrow0] (am) -- (recogniser);
    \draw[arrow0] (lm) -- (recogniser);
    \draw[arrow0] (fe) |- (am);
    
    \draw [dashed,gray] (-2.4,-0.9) rectangle (2.55,2.9);
    \draw [dashed,gray] (2.65,-0.9) rectangle (6.1,2.9);
    \draw [dashed,gray] (-2.5,-1) rectangle (6.2,3.5);

    \node [symbol,minimum height=0.3cm, text width=1.5cm, fill=gray!30!white] (frontEnd) at (0,2.7) {front-end};
    \node [symbol,minimum height=0.3cm, text width=1.5cm, fill=gray!30!white] (frontEnd) at (4.3,2.7) {back-end};
    \node [symbol,minimum height=0.3cm, text width=4.3cm, fill=gray!30!white] (frontEnd) at (2.0,3.3) {joint front- and back-end};

\end{tikzpicture}
}
